\documentclass[10pt,letterpaper,twocolumn]{article} %% two column, final layout

\usepackage{ol2}
\usepackage[draft]{hyperref}
\usepackage{amsmath}

\begin{document}

\twocolumn[ %% activate for two-column option

\title{Efficient generation of isolated attosecond pulses with high beam-quality by two-color Bessel-Gauss
beams }

\author{Zhe Wang$^1$, Weiyi Hong$^{1,3}$, Qingbin Zhang$^1$, Shaoyi Wang$^1$,  and Peixiang Lu$^{1,2,*}$}

\address{
$^1$Wuhan National Laboratory for Optoelectronics and School of
Physics, Huazhong University of Science and Technology,\\ Wuhan
430074, China
\\
$^2$School of Science, Wuhan Institute of Technology, Wuhan 430073, China \\

$^3$hongweiyi@mail.hust.edu.cn\\
$^*$Corresponding author: lupeixiang@mail.hust.edu.cn

}

\begin{abstract}The generation of isolated attosecond pulses with high efficiency and high beam
quality is essential for attosecond spectroscopy. We numerically
investigate the supercontinuum generation in a neutral rare-gas
medium driven by a two-color Bessel-Gauss beam. The results show
that an efficient smooth supercontinuum in the plateau is obtained
after propagation, and the spatial profile of the generated
attosecond pulse is Gaussian-like with the divergence angle of
$0.1^{\circ}$ in the far field. This bright source with high beam
quality is beneficial for detecting and controlling the
microscopic processes on attosecond time scale.
\end{abstract}

\ocis{190.7110, 190.4160, 300.6560.}
] %% activate for two-column option

\noindent The generation of isolated attosecond pulses makes it
possible to probe and control the ultrafast electronic processes
inside atoms with high time resolution, such as inner-shell
electronic relaxation or ionization by optical tunnelling
\cite{Kienberger}. So far, the most potential way to produce
isolated attosecond pulses in experiments is based on high-order
harmonic generation (HHG). Many efforts have been made on the
spectral and temporal characteristics of HHG in order to broaden
the bandwidth of the generated isolated attosecond pulses or
enhance the pulse efficiency \cite{Ferrari,Hong}. Moreover, the
spatial characteristics of the isolated attosecond source are also
crucial for attosecond science. The good beam quality of the
attosecond source leads to micrometer spot size and focused
intensities, which are required for applications in nonlinear
optics. The generation of efficient single attosecond pulse with
high beam quality is still challenging.

The isolated attosecond pulse generation is associated with the
electron dynamics in half a cycle of the driving field. The
three-step model \cite{Corkum}, which well depicts the HHG process
at the classical view, implies that HHG process can be controlled
via modulating the external driving field or the target to
generate a broadband supercontinuum. It has been proposed that a
two-color field \cite{Pfeifer,Cao} can control the acceleration
process to generate a broadband superconitnuum in the cutoff. Lan
\textit{et al}. \cite{Cao} also found that two-color scheme can
confine the ionization within half a cycle to produce efficient
supercontinuum in the plateau.

In the macro aspects of HHG process, the efficiency and the
spatial distribution of the generated attosecond pulse intensively
depend on the propagation and phase matching in the medium
\cite{Gaarde}. In a low-ionization medium, the two major factors
in phase matching are the intrinsic intensity-dependent phase
\cite{Philippe} and the geometrical phase induced by focusing of
the fundamental laser beam. Non-Gaussian or modified Gaussian
beams have been introduced to modulate the geometry of the driving
beams to optimize the phase matching condition for harmonics.
Bessel-Gauss(BG) beams offer a slowly varying geometry for
enhancement of the phase matching in HHG. T.Auguste \textit{et
al}. \cite{Auguste} reported a higher conversion efficiency and a
better spatial profile of the 21st-order harmonic with a BG beam
as compared with a Gaussian one in argon. Numerical studies
\cite{CRaron} and experiments \cite{Altucci} have been examined in
HHG, revealing that BG beams are more efficient than Gaussian
beams for generating harmonics in the plateau. In this letter, we
adopt a two-color BG beam as a pump pulse to investigate the
supercontinuum generation in helium. The results show that an
efficient isolated attosecond pulse is produced by the
supercontinuum in the plateau after propagation. Moreover, the
spatial profile of the generated attosecond pulse is Gaussian-like
with a small divergence angle in the far field. This efficient
attoseond pulse with high beam quality has important applications
in attosecond spectroscopy.

In this work, we perform a quantum simulation of helium in a
two-color intense laser field. The single-atom response is
calculated with Lewenstein model \cite{Lewenstein}, using the ADK
ionization rate \cite{Ammosov} and we solve the nonadiabatic 3D
light propagation for both fundamental and harmonic fields to
simulate the collective response of macroscopic gas. A 5 fs
linearly polarized fundamental pulse with a wavelength of 800 nm
and a 5 fs linearly polarized control pulse with a wavelength of
400 nm are used to synthesize the two-color driving field. The
peak intensities of the fundamental pulse and the control pulse
are $6\times10^{14}W/cm^2$ and $2.4\times10^{13}W/cm^2$,
respectively. The electric field of the two-color laser field is
given by
$E(r,z,t)=E_0(r,z)f(t)cos(w_0t)+E_1(r,z)f(t)cos(2w_0t+\varphi)$.
$E_0(r,z)$ and $E_1(r,z)$ are the amplitudes of the fundamental
field and the control field, respectively, which describe the
spatial distribution of the electric fields for BG beams. The
focusing half-angle of the two-color BG beam is set to
$0.5^{\circ}$. The envelope f(t) is a sine squared function and
$w_0$ is the frequency of the fundamental field. The relative
phase of the two fields $\varphi$ is set to zero to imitate the
ionization gating mechanism. The initial peak density of atoms is
$1.4\times10^{18}/cm^3$ and a 0.8 mm long gas jet with a truncated
Lorentzian density profile is placed 1 mm after the laser focus.

The spatiotemporal intensity profile of the synthesized driving
field is shown in Fig. 1 and the ionization rate of helium is
represented by the red line. For the two-color field, the
synthesized field can be modulated by mixing a control laser pulse
to the fundamental one. As shown in Fig. 1, the ionization of the
electron is enhanced at 2.5T and suppressed in other optical
cycles due to the shaping of the synthesized field. Therefore, HHG
is confined within half a cycle of the driving field and an
efficient supercontinuum can be produced in the first plateau,
which is called the the ionization gating scheme \cite{Cao}. The
spatial distribution of the intensity shows that the laser energy
is concentrated in the vicinity of the axis.

\begin{figure}[htb]
\centerline{\includegraphics[width=7.0cm]{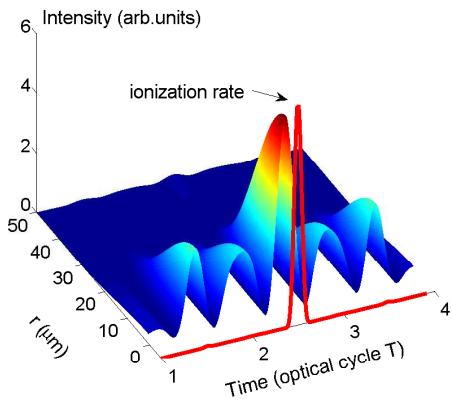}}\caption{
Spatiotemporal intensity profile of the driving laser pulse and
the ionization rate of helium.}
\end{figure}

\begin{figure}[htb]
\centering\includegraphics[width=8.0cm]{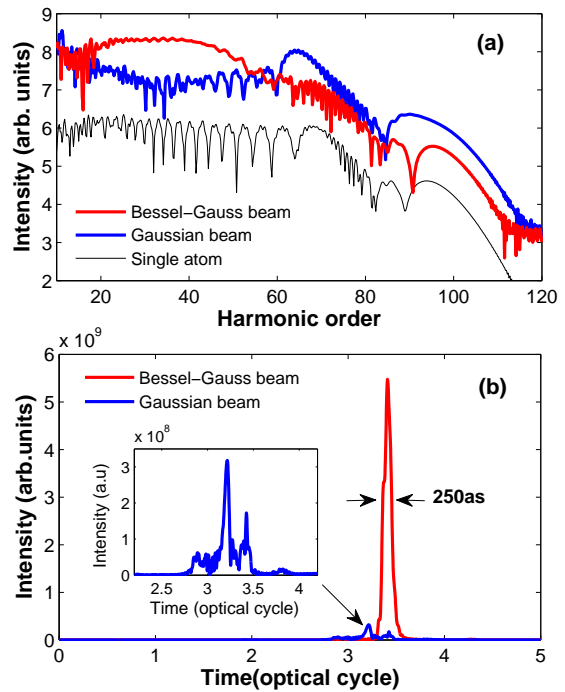} \caption{(a) The
harmonic spectrum and (b) temporal profiles of the attosecond
pulses by selecting 22-50 order harmonics. The inset is the
enlargement for the case of Gaussian beam.}
\end{figure}

The harmonic spectrum with the two-color BG beam after propagation
is presented by the red curve in Fig. 2(a). For comparison, the
harmonic spectrum with two-color associated Gaussian beam
\cite{Auguste} (blue curve) and the single-atom result (black
curve) are also presented, respectively. As shown in Fig. 2(a),
the harmonic spectrum for single-atom response in two-color field
is regularly modulated through the plateau to cutoff. The fringes
in the supercontinuum originate from the interference between two
quantum trajectories. After propagation, in the case of the
two-color Gaussian beam, there are still obvious modulations in
the supercontinuum. However, the interference in the
supercontinuum is largely removed and a smooth supercontinuum is
obtained in the first plateau (from 22th to 50th) for the
two-color BG beam case. Besides, the intensity of harmonics from
22th to 50th in the plateau with the two-color BG beam is enhanced
by about one order compared with the two-color Gaussian one. The
characteristics of the supercontinuum are associated with the
phase matching conditions in propagation. In our scheme, the
calculated ionization probability is below 0.1\%. Thus the two
governing factors in phase matching are the intrinsic
intensity-dependent phase and the geometrical phase of the driving
beam. Since the geometry of the BG beam is different from Gaussian
beam, phase matching conditions for the quantum paths (long and
short) differ between the two beams. In our simulation conditions,
in the case of two-color BG beam, for low-order harmonics, good
phase matching of only one path can be fully satisfied and a
single quantum path is macroscopically selected after propagation.
Thus the harmonics in the plateau are well phase matched and the
spectrum becomes continuous. As in the two-color Gaussian beam
case, both long path and short path survive after propagation. The
interference between quantum trajectories may limit the efficiency
of HHG and result in the modulations in the harmonic spectrum.
Fig. 2(b) shows the isolated attosecond pulse generated in our
scheme. By superposing the harmonics in the plateau (from 22th to
50th), a single 250as pulse is obtained using two-color BG beam.
The intensity of the attosecond pulse in the case of two-color
Gaussian beam is one order of magnitude lower and the inset in
Fig. 2(b) is the enlargement of its temporal profile. Moreover,
the temporal profile in two-color Gaussian beam case contains two
or more pulses, due to the poor phase matching for both short and
long quantum paths.

The far-field spatial characteristics of the attosecond source is
essential for its applications. We further investigate the spatial
profile of the isolated attosecond pulse in the far field through
a Hankel transformation \cite{Huillier}. The results are shown in
Fig. 3. For comparison, the far-field spatial profile of the
attosecond pulse with the two-color Gaussian beam is also
presented. Figure 3(a) and (b) are the spatial images of the
attosecond pulses with the two-color Gaussian and BG beams,
respectively. For the case of the Gaussian beam, the far-field
spatial distribution shows a annular-like structure. This implies
a large amount of the energy is radiated off axis and the
generated attosecond pulse is divergent, which may limit its
potential applications. For the case of the two-color BG beam, the
far-field spatial distribution shows a small size spot with
focused intensity and two annular rings with much lower
intensities. The divergence angle of central spot is approximately
$0.1^{\circ}$ from our calculation. The intensities of the two
additional annular rings are approximately 15\% and 10\% of the
central spot, respectively. These two rings can be filtered by a
aperture. Figure 3(c) and (d) present the far-field spatial
intensity profiles of the attosecond pulses with Gaussian and BG
beams, respectively. It is shown that the spatial intensity
profile with the two-color BG beam is Gaussian-like and its
intensity is nearly one order higher than that with the two-color
Gaussian beam. These spatial properties are related to the good
phase matching of the continuous harmonics near the axis. This
bright source with high beam quality is beneficial for many
potential applications, such as nonlinear studies and plasma
diagnostics.

\begin{figure}[htb]
\centering\includegraphics[width=8cm]{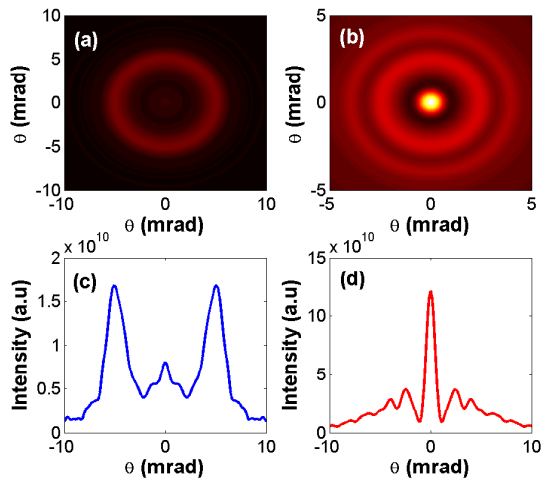} \caption{The far-field
profiles of isolated attosecond pulses with different driving
beams for the cases of two-color Gaussian beam (c) and two-color
Bessel-Gauss beam (d), and the corresponding spatial images of the
pulses are presented in (a) and (b). The far-field position is
located 1 m from the laser focus.}
\end{figure}

In summary, we have investigated the supercontinuum generation in
helium driven by a two-color BG beam based on the ionization
gating mechanism. It is shown that an efficient smooth
supercontinuum in the plateau (from 22th to 50th) is obtained
after propagation. In contrast to the conventional two-color
Gaussian beam, our calculations show that the beam quality of the
generated isolated attosecond pulse is significantly improved in
our scheme. The far-field spatial profile of the generated
attosecond pulse is Gaussian-like and the divergence angle is
approximately $0.1^{\circ}$. In addition, the intensity  of the
attosecond pulse generated is nearly one order higher than that
with the two-color Gaussian beam. This bright attosecond source
with high beam quality can  benefit many practical applications in
attosecond science. Experimentally, our scheme can be carried out
with a Ti:sapphire laser system. The laser beam is split into a
strong one and a weak one. The strong one is used as the
fundamental field. The weak one is used to produced the control
field via frequency multiplication technology. The driving beam
may be obtained by focusing a two-color Gaussian beam with an
axicon.

This work was supported by the National Natural Science Foundation
of China under Grants No. 60925021, 10734080, 11104092 and the 973
Program of China under Grant No. 2011CB808101.


\begin{thebibliography}{99}

\bibitem{Kienberger} R. Kienberger, E. Goulielmakis, M. Uiberacker, A. Baltuska, V.
Yakovlev, F. Bammer, A. Scrinzi, Th. Westerwalbesloh, U.
Kleineberg, U. Heinzmann, M. Drescher, and F. Krausz, Nature {\bf
427}, 817 (2004).

\bibitem{Ferrari} F. Ferrari, F. Calegari, M. Lucchini, C. Vozzi,
S. Stagira, G. Sansone, and M. Nisoli, Nat. Photon. {\bf 4}, 875
(2010).

\bibitem{Hong} W. Hong, P. Lu, Q. Li, and Q. Zhang, Opt. Lett. {\bf 34}, 2102 (2009).

\bibitem{Corkum} P. B. Corkum, Phys. Rev. Lett. {\bf 71}, 1994 (1993).

\bibitem{Pfeifer} T. Pfeifer, L. Gallmann, M. J. Abel, P. M. Nagel, D. M. Neumark,
and S. R. Leone, Phys. Rev. Lett. {\bf 97}, 163901 (2006).

\bibitem{Cao} P. F. Lan, P. X. Lu, W. Cao, Y. H. Li, and X. L. Wang, Phys. Rev.
A {\bf 76}, 051801(R) (2007).

\bibitem{Gaarde} M. B. Gaarde, J. L. Tate, and K. J. Schafer, J.
Phys. B: At. Mol. Opt. Phys. {\bf 41}, 132001 (2008).

\bibitem{Philippe} Philippe Balcou, Pascal Sali$\grave{e}$res, Anne
L'Huillier, and Maciej Lewenstein, Phys. Rev. A {\bf 55}, 3204
(1997).

\bibitem{Auguste} T. Auguste, O. Gobert, and B. Carr\'{e}, Phys. Rev. A {\bf 78}, 033411 (2008).

\bibitem{CRaron} C. F. R. Caron and R. M. Potvliege, Comput. Phys. Commun. {\bf 126}, 269 (2000).

\bibitem{Altucci} C. Altucci, R. Bruzzese, D. D'Antuoni, C. de
Lisio, and S. Solimeno, J. Opt. Soc. Am. B {\bf 17}, 34 (2000).

\bibitem{Lewenstein} M. Lewenstein, Ph. Balcou, M. Yu. Ivanov, A. L'Huillier, and P.
Corkum, Phys. Rev. A {\bf 49}, 2117 (1994).

\bibitem{Ammosov} M. V. Ammosov, N. B. Delone, and V. P. Krainov, Sov. Phys. JETP
{\bf 64}, 1191 (1986).

\bibitem{Huillier} A. L'Huillier, Ph. Balcou, S. Candel, K. J. Schafer, and K. C.
Kulander, Phys. Rev. A {\bf 46}, 2778 (1992).

\end{thebibliography}
\end{document}